\documentclass{article}

\usepackage{amssymb}
\usepackage{amsfonts}
\usepackage{amsmath}
\usepackage{graphicx}

\usepackage{hyperref}
\hypersetup{backref,
            pagebackref=true,
            ps2pdf,
            bookmarks=true,
            bookmarksnumbered=true,
            pdfauthor=McCabe Wydro,
            colorlinks=true,
            linkcolor=blue}

\title{Tests of Conformal Field Theory at the Yang-Lee Singularity}

\author{
\textrm{TOMASZ WYDRO}\\%
\textit{ Statistical Physics Group, P2M Dpt, Institut Jean Lamour}\\
  \textit{Nancy Universit\'e, Universit\'{e} Paul Verlaine - Metz}\\
  \textit{BP 70239, 54506 Vandoeuvre-les-Nancy Cedex, France}\\
wydro@lpm.u-nancy.fr
\\  \\
\textrm{JOHN F. McCABE}\\%
 \textit{ 2331 Gales Court, Scotch Plains, NJ 07076, USA}\\
jfmccabe@lycos.com
 }

\begin{document}

\maketitle

\begin{abstract}
This paper studies the Yang-Lee edge singularity of 2-dimensional
$(2D)$ Ising model based on a quantum spin chain and transfer matrix
measurements on the cylinder.  Based on finite-size scaling, the
low-lying excitation spectrum is found at the Yang-Lee edge
singularity. Based on transfer matrix techniques, the single
structure constant is evaluated at the Yang-Lee edge singularity.
The results of both types of measurements are found to be fully
consistent with the predictions for the $(A_{4}, A_{1})$ minimal
conformal field theory, which was previously identified with this
critical point.
\end{abstract}

 \maketitle

\section{1. Introduction}
In 1978, Fisher \cite{Fisher1} proposed that Yang-Lee edge
singularities \cite{YL1,YL2} are critical points. Later, Cardy
\cite{IsingYLS} argued that the Yang-Lee edge singularity of the
2D Ising model should be identified with the $(A_{4}, A_{1})$
minimal conformal field theory (CFT) \cite{BPZ,Friedan} of the ADE
classification \cite{ADE1}. Cardy's identification provides CFT
predictions for this Yang-Lee edge singularity.

This article tests different predictions coming from Cardy's
identification.

In section 2, we provide measurements of the low-lying excitation
spectrum at Yang-Lee edge singularity of the 2D Ising model. The
measured low-lying excitation spectrum is also compared with
predictions from Cardy's identification of the $(A_{4}, A_{1})$
minimal CFT with this Yang-Lee edge singularity of the 2D Ising
model \cite{IsingYLS,Zuber,Uzelac81}.

Cardy's identification also determines the forms of 2-point and
3-point correlations.  In particular, these correlations define
universal amplitudes, which are known as structure constants
\cite{BPZ,Dotsenko}.  Such predictions are an important advance that
CFT brought to the understanding of critical points of 2D
statistical models. No tests of such predictions have been performed
for critical points associated with non-unitary CFTs.

In section 3, we provide a measurement of the universal amplitude
associated with the Yang-Lee edge singularity of the 2D Ising model.
The measured amplitude is also compared with the prediction from
Cardy's identification of the $(A_{4}, A_{1})$ minimal CFT with this
Yang-Lee edge singularity.

\section{2. Excitation Spectrum at the Yang-Lee Edge Singularity
of the 2D Ising model}

The 2D Ising model in an imaginary external magnetic field is
associated with a quantum spin chain whose Hamiltonian,
$H_{Ising}$, on an $N$-site chain, is given by
\mbox{\cite{Gehlen87}}:
\begin{equation}
H_{Ising}=-\sum_{n=1}^{N}\{t\sigma_{z}(n)\sigma_{z}(n+1)+iB\sigma_{z}(n)+\sigma_{x}(n)\}{
.} \label{HIsing}
\end{equation}
In Eq. (\ref{HIsing}), $\sigma_{x}(n)$ and $\sigma_{z}(n)$ are
Pauli spin matrices at the site $n$, parameter "$t$" is a positive
coupling for a ferromagnetic spin-spin interaction, and $iB$ is a
purely imaginary external magnetic field.  In Eq. (\ref{HIsing}),
the last term produces inter-row single spin flips in the
associated 2D transfer matrix \mbox{\cite{Kogut,W-MC1}}.

Below, the phenomenological renormalization group (PRG) is used to
determine critical values of imaginary magnetic field,
$iB_{YL}(N)$, for various lengths, $N$, of the chain. For
imaginary magnetic fields, the PRG equation requires
that:\cite{Derrida,Zuber}
\begin{equation}
[N -1]m(B_{YL}(N), N-1)= [N]m(B_{YL}(N), N){ .} \label{PRG1}
\end{equation}
In Eq. (\ref{PRG1}), $m(B, N) = \left[E_{1}(B, N) - E_{0}(B,
N)\right]$ where $E_{0}(B, N)$ and $E_{1}(B,N)$ are energies for
the ground state "0" and the first excited state "1" on a chain of
length N. Below, $m(B, N)$ is referred to as $Gap(B, N)$ or more
simply as $Gap(N)$. At these $B_{YL}(N)$'s, the Ising quantum spin
chain exhibits the finite-size scaling behavior of the Yang-Lee
edge singularity. In particular, if the $B_{YL}(N)$'s converge to
a nonzero value as $N\rightarrow\infty$, that value will be the
critical point for the Yang-Lee edge singularity of the 2D Ising
spin model.

At these $B_{YL}(N)$'s, excitation energies should scale.  In
particular, CFT predicts how these energies will scale with the
length, $N$, of the chain. For an excited energy eigenstate "i" of
the quantum spin chain, an excitation energy, $E_{i}(N)-E_{0}(N)$,
will scale as:\cite{Cardy2}
\begin{equation}
E_{i}(N)-E_{0}(N)=\zeta 2\pi
\frac{\Delta_{i}+\bar{\Delta}_{i}-(\Delta+\bar{\Delta})}{N}{
.} \label{gaps scaling}
\end{equation}
In Eq. (\ref{gaps scaling}), $\Delta_{i}$ and $\bar{\Delta}_{i}$
are left and right conformal dimensions of conformal field "i",
and $\Delta$ and $\bar{\Delta}$ are conformal dimensions of the
primary field having the lowest "negative" scaling dimension in
the relevant non-unitary CFT. In Eq. (\ref{gaps scaling}), the
constant $\zeta$ is non-universal, e.g., depending on the
normalization of the Hamiltonian \footnote{$\zeta$ is the "sound
velocity" in the dispersion relation of the critical Hamiltonian}.

In minimal CFTs, the modular invariant forms of the partition
functions \cite{ADE1,RochaCarida} determine the low-lying
excitation spectrum and the central charges. For the $(A_{4},
A_{1})$ minimal CFT, Table \ref{tab:Spectra} gives the energies of
the low-lying excitations and degeneracies thereof as obtained
from the associated partition function.
\begin{table}[h]
\centering
\begin{tabular} {|c||c|c|c|c|c|c|c|c|c|}
\hline \multicolumn{1}{|c||}{CFT}&
\multicolumn{6}{c|}{$(A_{4},A_{1})$}\\
\hline
Normalized Energies &0&1&  2.5& 5.0& 6.0&   7.5\\
\hline
Degeneracy  &1&1&    2&    3&    2&     4\\
\hline
\end{tabular}
\caption{Lowest excitations of $(A_{4}, A_{1})$ CFT
\label{tab:Spectra}}
\end{table}
Table \ref{tab:Spectra} provides normalized excitation energies,
which are ratios. For a state "i", the normalized excitation energy
is the ratio is the excitation energy of the state "i" over the
excitation energy of the lowest excited state "1".  Here, excitation
energies are with respect to the ground state. The normalized
excitation energies of Table \ref{tab:Spectra} do not depend on
non-universal constants such as $\zeta$.

The critical magnetic fields, $B_{YL}(N)$, were obtained by
solving the PRG eq. (\ref{PRG1}) for chains of different lengths.
For these solutions, state energies were obtained by using the
Lanczos algorithm for $H_{Ising}$ of eq. (\ref{HIsing}). Table
\ref{tab: HC} shows critical fields, i.e., $B_{YL}(N)$'s, ground
state energies, and lowest excitation energies, i.e.,  $Gap(N)$'s.
These measurements were obtained for Ising quantum spin chains in
which the coupling, t, is 0.1. Table \ref{tab: HC} shows that
$NxGap(N)$ scales to a constant as $N\rightarrow\infty$ as
expected from the PRG.

\begin{table}[h]
{ \centering
\begin{tabular} {|c|c|c|c|c|}
\hline
 Number& & & & \\
of Sites &  $B_{YL}(N)$ & Energy of ground state & Gap(N) & $N \times Gap(N)$\\
\hline

3 & .2459180i  & -2.8811043 & .8103423  & 2.4310 \\
4 & .2384127i  & -3.8028211 & .6629112  & 2.6516 \\
5 & .2352339i  & -4.7341982 & .5613016  & 2.8065 \\
6 & .2337637i  & -5.6688215 & .4858628  & 2.9152 \\
7 & .2330279i  & -6.6048003 & .4275400  & 2.9928 \\
8 & .2326347i  & -7.5414746 & .3811698  & 3.0494 \\
9 & .2324118i  & -8.4785910 & .3435105  & 3.0916 \\
10& .2322793i  & -9.4160213 & .3123765  & 3.1237 \\
11& .2321972i  & -10.353696 & .286250   & 3.1488 \\
12& .2321442i  & -11.291568 & .264041   & 3.1685 \\
 ...   & ...   & ...        & ...       & ...    \\
$\infty$&.23193i&$-\infty$  & 0.0       & 3.2840 \\
\hline
\end{tabular}}
\caption{Measurements of $B_{YL}(N)$, Ground state energy, $Gap$,
and $NxGap$ for various chain lengths, $N$ \label{tab: HC}}
\end{table}

The critical magnetic field values, i.e., the $B_{YL}$'s, were
used to find the low-lying excitation spectra of Ising quantum
spin chains of various lengths. Table \ref{tab: mp} provides
measured spectra including both energies and degeneracies. Here,
excitation energies are also normalized by dividing by the lowest
excitation energy, i.e., as already described to remove any
dependence on the non-universal constant $\zeta$.

\begin{table}[t b p]
\centering
\begin{tabular} {|c||c|c|c|c|c|c|c|}
\hline
State /[Degeneracy] &  6    &  7     &  8   &   9    &     10   &    11   &  12  \\
 \hline
A / [2] & 2.68432  & 2.64386 & 2.61415 & 2.59207 & 2.57540 & 2.56260 & 2.55253 \\
\hline
B / [1] & 4.18193  & 4.27896 & 4.36713 & 4.44474 & 4.51197 & 4.56977 & 4.61912 \\
\hline
C / [2] & 4.51738  & 4.63236 & 4.70368 & 4.75182 & 4.78652 & 4.81281 & 4.83329 \\
\hline
D / [2] & 5.85889  & 5.89208 & 5.91240 & 5.92644 & 5.93703 & 5.94544 & 5.95210 \\
\hline
E / [2] &    --    & 5.68559 & 6.03104 & 6.27270 & 6.45018 & 6.58573 & 6.69223 \\
\hline
F / [2] &    --    & 6.24252 & 6.35798 & 6.46344 & 6.55966 & 6.64694 & 6.72535 \\
\hline
\end{tabular}
\caption{Normalized excitation energies and degeneracies of lowest
excited states A - F for Ising quantum spin chains with 6 to 12
sites. \label{tab: mp}}
\end{table}

Figures 1 - 4 plot the measured excitation energies of the states
A - F as a function of the inverse of the length of the Ising
quantum spin chain.

\begin{figure}[h t p ]
    \includegraphics{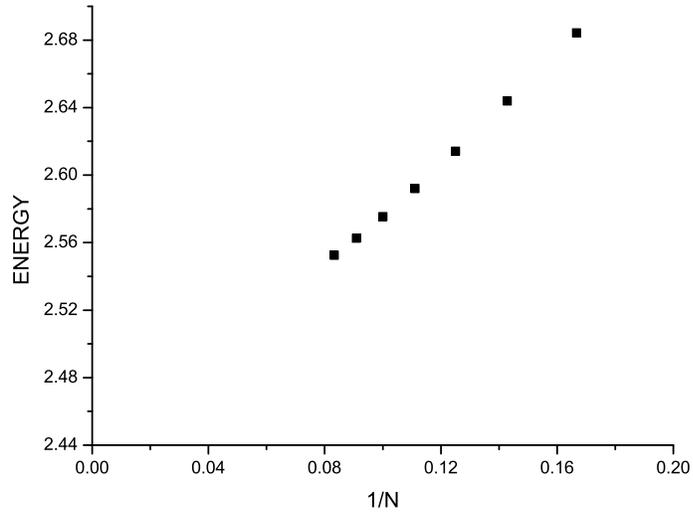}%
    \caption{Energies of the type A states as a function of $1/N$.}%
    \label{fig:A}
\end{figure}

\begin{figure}[h t p]
    \includegraphics{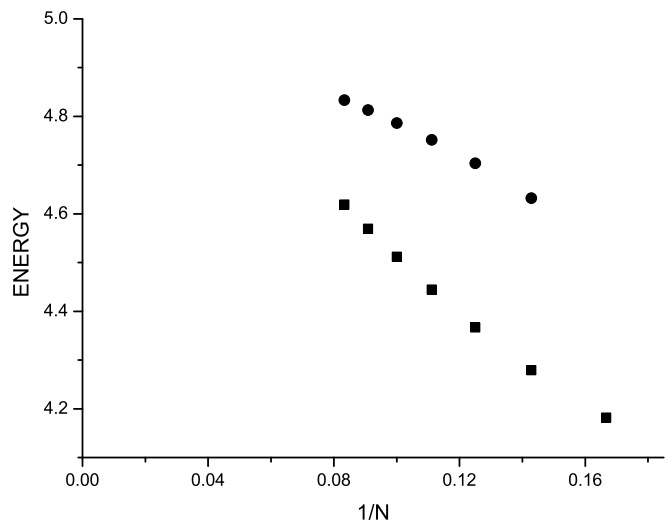}%
    \caption{Energies of type B states (squares) and type C states (circles) as a function of $1/N$.}%
    \label{fig:BandC}
\end{figure}

\begin{figure}[h t p]
    \includegraphics{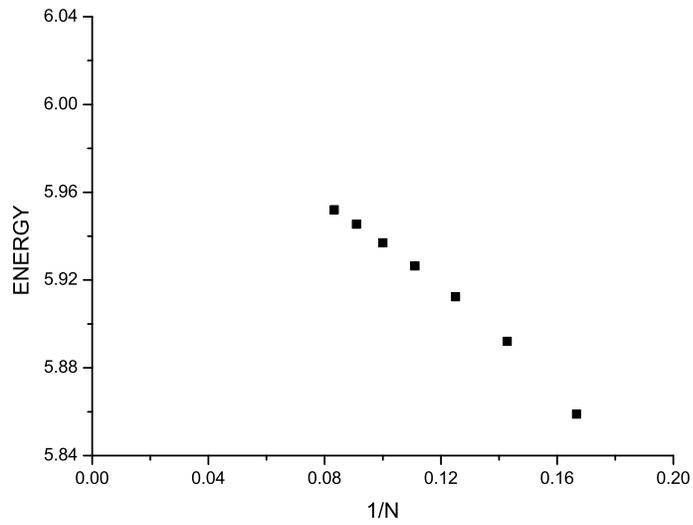}%
    \caption{Energies of type D states as a function of $1/N$.}%
    \label{fig:D}
\end{figure}

\begin{figure}[h t p]
    \includegraphics{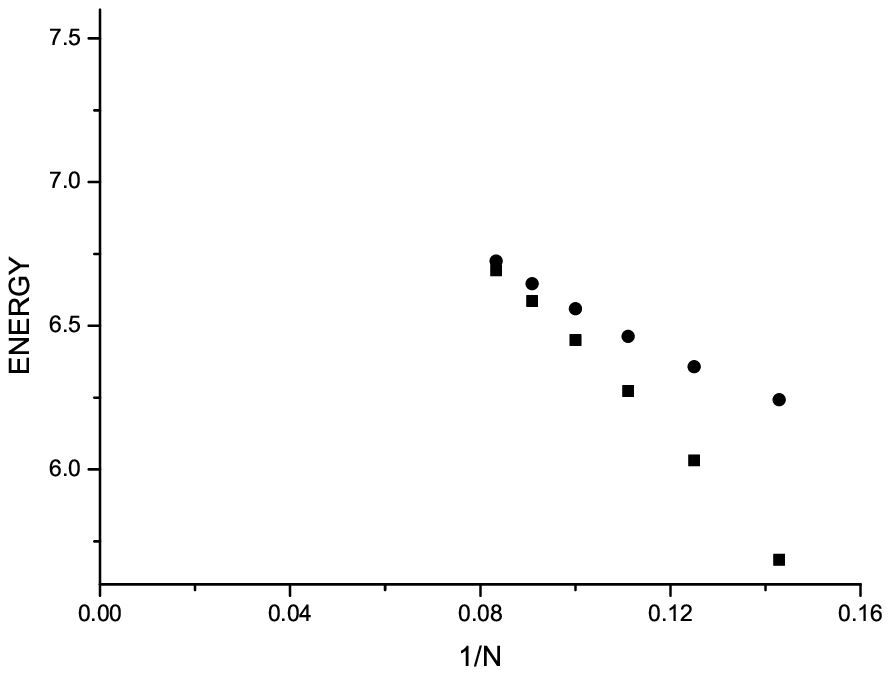}%
    \caption{Energies of type E states (squares) and F states (circles) as a function of $1/N$.}%
    \label{fig:EandF}
\end{figure}

A visual inspection of Figures 1 - 4 shows that the type A, B, C,
D, E, and F states form four distinct sets A, B $\&$ C, D, and E
$\&$ F. Within each set, the states have energies that approach
the same value as $1/N \rightarrow 0$.  The excitation energies of
the states of sets A, B $\&$ C, D, and E $\&$ F approach about
2.45, 5.0, 6.03, and 7.6, respectively, as $1/N \rightarrow 0$.  A
BST analysis shows that excitation energies of the type A, B, C,
D, E, and F states scale to 2.4995(5), 5.005(1), 5.003(3),
5.99(1), 7.54(8), and 7.60(7), respectively, in this limit. These
PRG measurements of the low-lying excitation energies and
degeneracies agree well with the predictions for the $(A_4, A_1)$
CFT as in Table \ref{tab:Spectra}.

\section{3. Structure Constant at the Yang-Lee Edge Singularity}
The non-unitary $(A_{4}, A_{1})$ minimal CFT has one primary field
$\phi(z,\bar z)$ with left and right conformal weights -1/5 and
scaling dimension $x$ of -2/5 \cite{IsingYLS}. For this field,
$\phi(z,\bar z)$, 2-point and 3-point correlations have the forms:
\begin{equation}
G_{\phi\phi}(z_1, \bar{z_1}, z_2, \bar{z_2}) = |(z_1-z_2)|^{4/5},
\label{2point}
\end{equation}and
\begin{equation}
G_{\phi\phi\phi}(z_1, \bar{z_1}, z_2, \bar{z_2}, z_3, \bar{z_3}) =
C|(z_1-z_2)(z_2-z_3)(z_3-z_1)|^{2/5}. \label{3point}
\end{equation}
Cardy showed that the structure constant, $C$, of the non-unitary
$(A_{4}, A_{1})$ minimal CFT is given by:\cite{IsingYLS}
\begin{equation}
C=\sqrt{-\frac{[\Gamma(6/5)]^2\Gamma(1/5)\Gamma(2/5)}{\Gamma(3/5)[\Gamma(4/5)]^3}}
\label{eq:StructConstExact}
\end{equation}
Below, numerical measurements at the Yang-Lee edge singularity are
presented for this CFT prediction. The numerical measurements were
made for the 2D Ising model, i.e., rather than for a spin chain. The
2D Ising model has a Hamiltonian $H$, given by:
\begin{equation}
H=-\sum_{j=1}^{M}\sum_{i=1}^{N}[J(S_{i,j}S_{i,j+1}+S_{i,j}S_{i+1,j})+hS_{i,j}]
. \label{eq:IsingHamilt}
\end{equation}
The spin-spin coupling $J$ is positive. In this model, the
Yang-Lee edge singularity occurs above the critical temperature
for a purely imaginary values of the magnetic field, $h$, i.e., $h
= iB$ with $B$ real \cite{YL1,YL2}. Below, the spin correlations
were measured at a temperature, $T$, for which $J/k_BT = 0.1$

The transfer matrix was used to measure correlation 2-spin and
3-spin correlations on torii of length, $M$, and diameters, $N$. In
these evaluations, $M$ was much larger than $N$, i.e., M = 512 and N
= 3 - 8, so that correlations had distance behaviors for infinitely
long cylinders at field separations small compared to $M$.

Finite-size scaling enabled the extraction of physical properties in
the thermodynamic limit \cite{FiniteScall}. In particular, the spin
correlations were measured at purely imaginary magnetic field
values, $h(N) = iB_{YL}(N)$.  Each value , $B_{YL}(N)$, satisfied
the phenomenological renormalization group (PRG) equation for
infinite cylinders of diameters $(N-1)$ and $N$:
\begin{equation}
\frac{\xi(iB_{YL}(N), N-1)}{N-1}=\frac{ \xi(iB_{YL}(N),
N)}{N}{ .} \label{PRG}
\end{equation}
In the PRG equation, $\xi(iB, N)$ is the spin-spin correlation
length on the infinite cylinder of \mbox{diameter} $N$ at the
magnetic field $iB$ \footnote{$\xi$ is measured by the first
inverse gap (see \cite{Kogut})}.

On a cylinder of width $N$, CFT predicts that correlations depend
exponentially on distances between fields when said distances are
large compared the cylinder's diameter, $N$ \cite{Cardy2}. When
$|y_1 - y_2| >> N$, the 2-point correlation of fields of scaling
dimension, $x$, has the form $\exp(-2\pi x (y_1 - y_2)/N)$ where
$y_1$ and $y_2$ are the positions of the fields along the axis of
the infinite cylinder. For the 3-point correlation, the exponential
behavior on the distances between the fields of the correlation is
also determined by the scaling dimensions of the fields therein

At the Yang-Lee edge singularity, amplitudes of 2-spin and 3-spin
correlations, i.e., $A_{ss}$ and $A_{sss}$, were used to evaluate
the 3-spin structure constant. The values of the 3-spin structure
constant, $C(N)$, were obtained from the relation:
\begin{equation}
C(N)=\frac{A_{sss}(iB_{YL}(N))}{[A_{ss}(iB_{YL}(N)]^{3/2}}.
\end{equation}
In the above equation, $A_{ss}(iB_{YL}(N))$ and
$A_{sss}(iB_{YL}(N))$, are amplitudes of the respective 2-spin and
3-spin correlations at PRG values for the magnetic field. PRG
measurements of the correlation length $\xi(N)$ also provide a
measurement of the conformal dimension, $x$, of the spin field,
i.e., $x(N) = N/[2\pi\xi(iB_{YL}(N))]$. Scaling behaviors of these
quantities with the cylinder's width, $N$, were used to obtain the
values of the quantities as $N\to\infty$.

Table \ref{tab:CriticalFields} provides our transfer matrix
measurements\footnote{For N = 3 - 7, $B_{YL}(N)$'s were evaluated
from the PRG equation. For N = 8, $B_{YL}(8)$' was estimated from
the $B_{YL}(N)$'s for N = 3 - 7 by assuming a leading finite-size
scaling behavior.} for $M = 512$ and $J/k_BT = 0.1$.
\begin{table}[h]
\vspace{1mm} \centering
\begin{tabular} {clll}
\hline
 $N$ &  $B_{YL}(N)$ & $x(N)$ & $|C(N)|$ \\
\hline
3 & 0.184802 & 0.353929 & 1.80838\\
4 & 0.183348 & 0.376870 & 1.83711\\
5 & 0.183064 & 0.385748 & 1.85736\\
6 & 0.182982 & 0.390108 & 1.87054\\
7 & 0.182951 & 0.392693 & 1.87937\\
8 & 0.182946 & 0.392911 & 1.88633\\
\ldots&\ldots&\ldots&\ldots\\
$\infty$&\ldots&0.398(2)&1.923(13)\\
$CFT$& --- &0.4&1.9113 \\
\hline
\end{tabular}
\caption{PRG Measured values of conformal dimension and structure
constant. \label{tab:CriticalFields}}
\end{table}

In Table \ref{tab:CriticalFields}, the $\infty$ line extrapolates
the measured values to the thermodynamic limit, i.e., $N = \infty$.
The extrapolated values were obtained from fits of the measured
$x(N)$'s and $|C(N)|$'s to functions of form $f(N) = f(\infty)+ f_1
N^{-\alpha} $, i.e., leading finite-size scaling forms.

In Table \ref{tab:CriticalFields}, the last line gives the
predictions for $x$ and $|C|$ from the $(A_{4}, A_{1})$ non-unitary
minimal CFT model.

\begin{figure} [h t p]
 {\centering\includegraphics[scale=1.2]{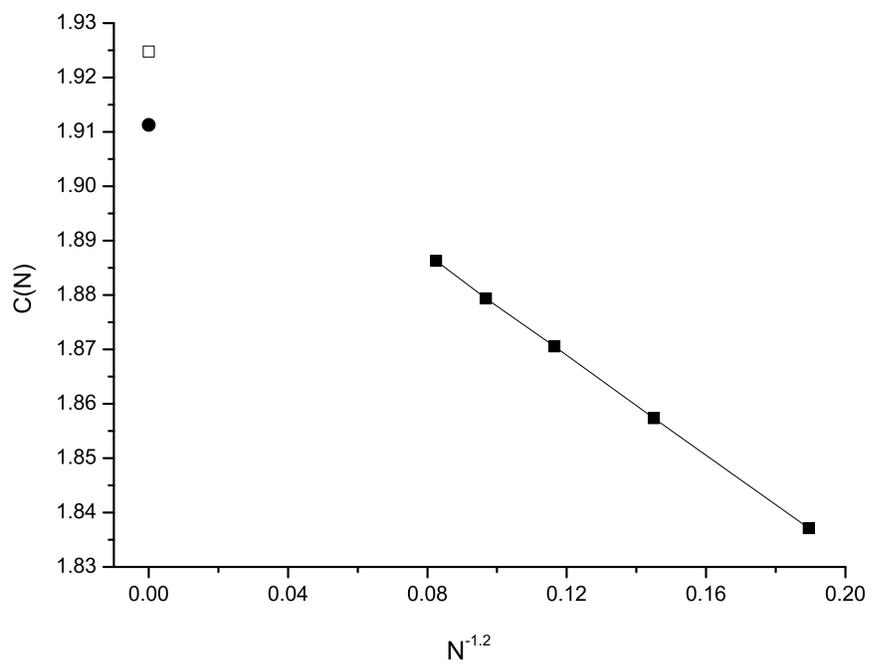}}
\caption{The measured structure constant (squares) for cylinders
of diameter $N$ plotted against nonlinear fit for $N= 4 - 8.$
\label{fig:StructConstScaling}}
\end{figure}

Figure \ref{fig:StructConstScaling} shows measurements of the
structure constant $|C(N)|$ and a best fit (line), which accounts
for a correction in $N^{-1.2}$. In Figure
\ref{fig:StructConstScaling}, the black squares are the measured
$C(N)$'s, the empty square is the value of $C(\infty)$ from the
best fit, and the black circle is the CFT prediction.  Our
finite-size scaling measurements produce a value for the structure
constant that again agrees well with the prediction of the $(A_4,
A_1)$ minimal CFT.


\bigskip

\newpage

\begin{thebibliography}{99}

\bibitem{Fisher1} M.E. Fisher, Phys. Rev. Lett. 40 (1978) 1610; see also
D.A. Kurtz and M.E. Fisher, Journ. of Stat. Phys. 19 (1978) 205;
Phys. Rev. B 20 (1979) 2785.

\bibitem{YL1} C.N. Yang, T.D. Lee, Phys. Rev. 87 (1952) 404.

\bibitem{YL2} C.N. Yang, T.D. Lee, Phys. Rev. 87 (1952) 410.

\bibitem{IsingYLS} J.L. Cardy, Phys. Rev. Lett. {\bf 54}, 1354 (1985).

\bibitem{BPZ} A.A. Belavin, A.M. Polyakov, and A.B. Zamolodchikov, Nucl.
Phys. B 241 (1984) 333; for a review see, e.g.: M. Henkel,
\underline{Conformal Invariance and Critical Phenomena}
(Springer-Verlag, Germany 1999); P. Di Francesco, P. Mathieu, and
D. S\'{e}n\'{e}chal, \underline{Conformal Field Theory}
(Springer-Verlag, USA 1997) pages 210 - 228 .

\bibitem{Friedan} D. Friedan, Z. Qiu, and S. Shenker, Phys. Rev. Lett. {\bf 52}, 1575
(1984), and Comm. Math. Phys. {\bf 107}, 535 (1986).

\bibitem{ADE1} A. Cappelli, C. Itzykson, J.-B. Zuber, Nucl. Phys. B 280 (1987)
445; Comm. Math. Phys. 113 (1987) 1; A. Kato, Mod. Phys. Lett. A 2
(1987) 585; for a review see e.g., C. Itzykson and J.-M. Drouffe,
\underline{Statistical Field Theory} (Cambridge University Press,
U.K. 1989) chapter IX (1989).

\bibitem{Zuber} C. Itzykson, H. Saleur, and J.-B. Zuber, Europhys. Lett. 2
(1986) 91.

\bibitem{Uzelac81} K. Uzelac and R. Jullien, J. Phys. A 14 (1981) L151.

\bibitem{Dotsenko} Vl.S. Dotsenko and V.A. Fateev, Nucl. Phys. {\bf B240}, 312 (1984),
and {\bf B251}, 691 (1985).

\bibitem{Derrida} M.P. Nightingale, Physica 83 A (1976) 561; B. Derrida
and L. de Seze, J. Physique 43 (1982) 475; see, e.g., M.P.
Nightingale, in {\underline Finite Size Scaling and Numerical
Simulation of Statistical Systems} (World Scientific Publishing,
Singapore, Ed. V. Privman 1990) Ch. VII.

\bibitem{RochaCarida} A. Rocha-Caridi in \underline{Vertex Operators in Mathematics and
Physics} MSRI Publications No. 3 (Springer, USA, Eds. J. Lepowski,
S. Mandelstam ad I.M. Singer 1985) 451.

\bibitem{FiniteScall} M.E. Fisher and M.N. Barber, Phys. Rev. Lett. 28
(1972) 1516; for a brief review see, e.g., M. Henkel,
\underline{Conformal Invariance and Critical Phenomena}
(Springer-Verlag, Germany 1999) chapter 3.

\bibitem{Cardy2} J.L. Cardy, J. Phys. {\bf A17}, L385 (1984), and
Nucl. Phys. {\bf B270}, 186 (1986); for a review see M. Henkel,
{\it Conformal Invariance and Critical Phenomena}
(Springer-Verlag, Germany, 1999) Ch. 13.

\bibitem{Gehlen87} M. Suzuki, Prog. Theor. Phys. 56 (1976) 1454;
G. von Gehlen, V. Rittenberg, and T. Vescan, J. Phys. A 20 (1987)
2577; for a review see e.g., M. Henkel in \underline{Conformal
Invariance and Critical Phenomena} (Springer-Verlag, Germany 1999)
Chs. 8-10.

\bibitem{Kogut}J.B. Kogut, Rev. Mod. Phys. 51 (1979) 659.

\bibitem{W-MC1} See, e.g., the review of T. Wydro and J. McCabe,
in \underline{Proceedings} \underline{ of the 7th International
School on Theoretical Physics "Symmetry and Structural}
\underline{ Properties of Condensed Matter"} (World Scientific
Publishing, Singapore, Eds. T. Lulek, B. Lulek, and A. Wal
Singapore 2003) 9.

\bibitem{W-MC2}T. Wydro and J. McCabe, Int. J. of Mod. Phys. B 19 (2005)
3021.



\end{thebibliography}
\end{document}